\documentclass{PoS}
\usepackage{amsmath,epsfig,cite,graphics}
\usepackage{macros_static,macros}
\usepackage{wrapfig}

\title{Efficient use of the Generalized Eigenvalue Problem}

\ShortTitle{Efficient use of the Generalized Eigenvalue Problem}

\author{\LALPHA \hfill
        \onecol{4.0cm}{\vspace{-1.5cm}\it DESY 08-108 \\ SFB/CPP-08-55
          \\CERN-PH-TH/2008-168 
          \vspace{-1.7cm}
        }}
\author{B. Blossier, G. von Hippel, T. Mendes\thanks{
        permanent address: {\em IFSC, University of S\~ao Paulo, 
        C.P. 369, CEP 13560-970, S\~ao Carlos SP, Brazil}.
        }, \speaker{R. Sommer}
        \\
        DESY, Platanenalle 6, 15738 Zeuthen, Germany\\
        E-mail: \email{rainer.sommer@desy.de}
        }
\author{M. Della Morte\\
        CERN, Physics Department, TH Unit,
        CH-1211 Geneva~23, Switzerland
       }

\abstract{
We analyze the systematic errors made when using the
generalized eigenvalue problem to extract energies
and matrix elements in lattice gauge theory. Effective
theories such as HQET are also discussed.
Numerical results are shown for the extraction of
ground-state and excited B-meson masses and the 
ground-state decay constant in the static approximation.
}

\FullConference{The XXVI International Symposium on Lattice Field Theory\\
		July 14--19 2008 \\
		Williamsburg, Virginia, USA}

\begin{document}

\section{The Generalized Eigenvalue Problem \label{s:gevp}}
\subsection{History}
At a conference in 1981, K. Wilson suggested to use
a variational technique to compute energy levels
in lattice gauge theory \cite{Wilson}. The idea
was picked up and applied to the glueball
spectrum\cite{gevp:berg,gevp:michael} and to the static quark
potential(s)\cite{gevp:pot}. With a certain choice of the 
variational basis $\left\{\phi_i\,,\,i=1\ldots N\right\}$ and maximizing 
$\langle \phi |\rme^{-(t-t_0)\hat H} | \phi \rangle/
\langle \phi | \phi \rangle$ with 
$|\phi\rangle = \sum_i \alpha_i |\phi_i\rangle$,
the variational technique yields the generalized eigenvalue problem (GEVP).
It is applicable beyond the computation of the ground-state energy
and has been widely used, but rarely in the form where
it can be shown that corrections to the true energy levels
decrease exponentially for large time \cite{phaseshifts:LW}. 

Apart from \cite{phaseshifts:LW}, statements about corrections due to higher 
energy levels seem to be absent in the literature. We here add such statements
and suggest a somewhat different use of the GEVP, which we will show
    to be more efficient under certain conditions. We also treat the case of
    an 
effective theory and show 
numerical results for heavy-quark effective theory (HQET).
 
\subsection{Basic idea}

We start from a matrix of correlation functions on an infinite-time 
lattice
\bes
  \label{e:cij}
  C_{ij}(t) &=& \langle O_i(0) O_j(t) \rangle =
  \sum_{n=1}^\infty \rme^{-E_n t} \psi_{ni}\psi_{nj}\,,\quad 
  i,j = 1,\ldots,N  \\ && \quad \psi_{ni} \equiv (\psi_n)_i = 
  \langle n|\hat O_i|0\rangle
  \quad
  E_n \leq E_{n+1} \,.
 \nonumber 
\ees 
For simplicity we assume real $\psi_{ni}$. States $|n\rangle$ with
$\langle m|n\rangle =\delta_{mn}$ are eigenstates of the transfer matrix
and all energies have the vacuum energy subtracted.
$O_j(t)$ are any gauge-invariant fields on a timeslice $t$ that correspond
to Hilbert-space operators $\hat O_j$ whose quantum numbers are
then also carried by the states $|n\rangle$. 
Besides the energy levels $E_n$ one may want to determine a
matrix element 
\bes
    p_{0n} = \langle 0|\hat P|n\rangle
\ees
of an operator $\hat P$ that may or may not be in the set of
operators $\left\{ \hat O_i \right\}$. 
Starting from the GEVP, 
\bes \label{e:gevp}
  C(t)\, v_n(t,t_0) = \lambda_n(t,t_0)\, C(t_0)\,v_n(t,t_0) \,, 
  \quad n=1,\ldots,N\,\quad t>t_0,   
\ees
L\"uscher and Wolff showed that \cite{phaseshifts:LW} 
\bes
  E_n = \lim_{t\to\infty} E_n^\mrm{eff}(t,t_0)\,,
  \quad E_n^\mrm{eff}(t,t_0)= {1\over a} \, \log{\lambda_n(t,t_0) \over
    \lambda_n(t+a,t_0)}\,.
\ees
For a while we now assume that only $N$ states contribute,
\bes
  C_{ij}(t)= C_{ij}^{(0)}(t) = \sum_{n=1}^N \rme^{-E_n t} \psi_{ni}\psi_{nj}
                  \,. \label{e:C0}
\ees
We introduce the dual (time-independent) vectors $u_n$, defined by
$(u_n,\psi_m) = \delta_{mn}\,,\; m,n\leq N\,$,
with $(u_n,\psi_m)\equiv \sum_{i=1}^N (u_n)_i\,\psi_{mi}$. Inserting into
\eq{e:C0} gives
\bes
  C^{(0)}(t) u_n = \rme^{-E_n t} \psi_n \,, \label{e:vn0}\quad
  C^{(0)}(t)\, u_n = \lambda_n^{(0)}(t,t_0)\, C^{(0)}(t_0)\,u_n\,.
                                \label{eq:gevp0}
\ees
So the GEVP is solved by
\bes
  \lambda_n^{(0)}(t,t_0) = \rme^{-E_n (t-t_0)}\,, \quad
  v_n(t,t_0) \propto u_n
\label{eq:lambda0}
\ees
and there is an orthogonality for all $t$ of the form
\bes
  (u_m,C^{(0)}(t)\,u_n) = \delta_{mn}\,\rho_n(t)\,,
  \quad \rho_n(t) = \rme^{-E_n t}\,.
  \label{e:norm}
\ees
These equations mean that the operators
$
  \hat Q_n = \sum_{i=1}^N (u_n)_i \hat O_i \equiv (\hat O,u_n)\,
$
create the eigenstates 
$\,|n\rangle = \hat Q_n|0 \rangle \,$
of the Hamilton operator:
$
\hat H|n \rangle =E_n\,|n \rangle\,.
$
Consequently we have 
$
    p_{0n} = \langle 0|\hat P|n\rangle =  \langle 0| \hat P \hat Q_n
    |0\rangle 
$,
which, preparing for a generalization,
we may rewrite as
\bes
    p_{0n} = \sum_{j=1}^N\, \langle P(t) O_j(0) \rangle (u_n)_j 
    =  {  \sum_{j=1}^N\,\langle P(t) O_j(0)\rangle \,v_n(t,t_0)_j \over 
               \left(v_n(t,t_0)\,,\, C(t)\,v_n(t,t_0)\right)^{1/2}}
               {\lambda_n(t_0+t/2,t_0) \over \lambda_n(t_0+t,t_0)}\,,
\ees
while for all $t,t_0$ we have
$
    E_n^\mrm{eff}(t,t_0) = E_n \,.
$

Let us now come back to the general case \eq{e:cij}. The idea is to solve
the GEVP, \eq{e:gevp}, ``at large time'' where the contribution of states $n>N$ is
small and obtain matrix elements and energy levels from
\newcommand{\corren}{\varepsilon_{n}}
\newcommand{\corrpn}{\pi_{n}}
\bes
  \label{e:eneff}
  E_n^\mrm{eff} &=& {1\over a} \, \log{\lambda_n(t,t_0) \over
    \lambda_n(t+a,t_0)}\, \;=\; E_n + \corren(t,t_0) \\[1ex]
    p_{0n}^\mrm{eff} &=& { \sum_{j=1}^N\,\langle P(t_1) O_j(0)\rangle \,(v_n(t,t_0))_j \over 
               \left(v_n(t,t_0)\,,\, C(t_2)\,v_n(t,t_0)\right)^{1/2}}
               {\lambda_n(t_0+t_2/2,t_0) \over \lambda_n(t_0+t_1,t_0)}  
           \;=\; p_{0n}
               + \corrpn(t,t_0)\quad\mbox{at\ }t_1=t_2=t \,.\qquad
\ees
The restriction to $t_1=t_2=t$ is for simplicity.
The corrections $\corren,\corrpn$ will disappear at large times.
Note that in the literature
the energy levels are often not extracted in this way. Rather, the standard 
effective masses of correlators made from $Q_n=(O,v_n(t,t_0))$ are used, 
and the question of the size of the corrections is left open.
However, the form in \eq{e:eneff} has a theoretical advantage as it
was shown in
 \cite{phaseshifts:LW} that (at fixed $t_0$)
\bes
  \corren(t,t_0) = \rmO(\rme^{-\Delta E_n\, t})\,,\quad \Delta E_n =
  \min_{m\neq n}\, |E_m-E_n|\,.
\ees
This is non-trivial as it allows to obtain the excited levels
with corrections that vanish in the limit of large $t$, keeping $t_0$ fixed.
However, it appears from this formula that the corrections can be very large
when there is an energy level close to the desired one. This is the case
in interesting phenomena such as string breaking 
\cite{pot:Higgs1,pot:Higgs2}, where in numerical 
applications the
corrections appeared to be very small despite the formula 
above\footnote{In fact a different formula was claimed in \cite{pot:Higgs1}.}.
Also in static-light systems the gaps
are typically only around $\Delta E_n \approx 400\,\MeV$,
and in full QCD with light quarks a small gap 
 $\Delta E_n \approx 2 m_\pi$ appears in some channels.

Our contribution to the issue is a more complete discussion of the 
correction $\corren$ to $E_n$ as well as a discussion of the 
corrections $\corrpn$ to the
matrix elements. It turns out that a very useful case is to consider
the situation 
\bes
  t \leq 2 t_0\,,
\ees
e.g.\ with $t-t_0 = \mbox{const.}$ or $2\geq t/t_0=\mbox{const.}$,
and then take $t_0$ (in practice moderately) large. Then it is not difficult
to show that 
\bes
  \label{e:corren}
  \corren(t,t_0) &=& \rmO(\rme^{-\Delta E_{N+1,n}\, t}) \,,\quad 
                   \Delta E_{m,n} = E_m-E_n \,, \\
  \label{e:corrpn}
  \corrpn(t,t_0) &=& \rmO(\rme^{-\Delta E_{N+1,n}\, t_0})\,,\quad \mbox{at
    fixed } t-t_0 \\
  \label{e:corrp1}
  \pi_1(t,t_0) &=& \rmO(\rme^{-\Delta E_{N+1,1}\, t_0} 
  \rme^{-\Delta E_{2,1}\,(t- t_0)}) \,+\, \rmO(\rme^{-\Delta E_{N+1,1}\, t}) 
   \,.
\ees
The large gaps $\Delta E_{N+1,n}$ can solve the problem of close-by levels
for example in the string-breaking situation, but also speed
up the general convergence very much. For example in static-light systems
$\Delta E_{6,1} \approx 2\,\GeV$ means that roughly a factor of 5 in time
separation is gained.  We now turn to an outline of the proof of these
statements.

\section{Perturbation theory \label{s:pt}}
We start from the solutions above
for $C=C^{(0)}$ and treat the higher states as perturbations. 
This perturbative evaluation was 
already set up by F. Niedermayer and P. Weisz a while 
ago \cite{notes:FerencPeter} but never published. We noted 
the advantage of $ t \leq 2 t_0$, the form of the corrections
to the effective matrix elements defined above and could show
that these relations hold to all orders in the
expansion.    

We want to obtain $\lambda_n$ and ${v}_n$ in a perturbation
theory in $\eps$, where
\bes
  A{v}_n = \lambda_n B {v}_n\, ,\quad A=A^{(0)}+\eps A^{(1)} \, ,\quad
  B=B^{(0)}+\eps B^{(1)} \,.
\ees
We will set
\bea
  \label{e:A}
  A^{(0)}&=& C^{(0)}(t)\,,\quad \eps A^{(1)}= C^{(1)}(t) \,,\\ 
  \label{e:B}
  B^{(0)}&=& C^{(0)}(t_0)\,,\quad \eps B^{(1)}= C^{(1)}(t_0) \,
\ees
in the end. The solutions of the lowest-order equation
$
  A^{(0)}{v}_n^{(0)} \;=\; \lambda_n^{(0)} B^{(0)}\, {v}_n^{(0)}
$
satisfy an orthogonality relation 
$
\,  ({v}_n^{(0)},B^{(0)}{v}_m^{(0)}) = \rho_n \, \delta_{nm}\,
  \label{eq:rho}
$
 as in \eq{e:norm} above.
Writing 
\bes
  \lambda_n \;=\; \lambda_n^{(0)}+\eps \lambda_n^{(1)} +\eps^2 \lambda_n^{(2)}\,\ldots 
  \,,\quad
  {v}_n \;=\; {v}_n^{(0)}+\eps {v}_n^{(1)} +\eps^2 {v}_n^{(2)}\,\ldots 
\ees
we get for the first two orders
\bes
A^{(0)} {v}_n^{(1)} + A^{(1)} {v}_n^{(0)}  &=&
\lambda_n^{(0)}\,\left[B^{(0)} {v}_n^{(1)} + B^{(1)} {v}_n^{(0)}\right]
+\lambda_n^{(1)}\,B^{(0)} {v}_n^{(0)} \,, \label{eq:1stord} \\[1mm]
A^{(0)} {v}_n^{(2)} + A^{(1)} {v}_n^{(1)}  &=&
\lambda_n^{(0)}\,\left[B^{(0)} {v}_n^{(2)} + B^{(1)} {v}_n^{(1)}\right]
+\lambda_n^{(1)}\,\left[B^{(0)} {v}_n^{(1)} + B^{(1)} {v}_n^{(0)}\right]
+\lambda_n^{(2)}\,B^{(0)} {v}_n^{(0)} \,. \quad \label{eq:2ndord} \quad
\ees

With the orthogonality of the lowest-order vectors, ${v}_n^{(0)}$, one obtains 
just like in 
ordinary QM perturbation theory the solutions for eigenvalues and
eigenvectors
\bes
  \lambda_n^{(1)} &=&  \rho_n^{-1}\,
                  \left({v}_n^{(0)},\Delta_n {v}_n^{(0)}\right)\,,
                  \quad \Delta_n \,\equiv\, A^{(1)} - \lambda_n^{(0)} B^{(1)}\, \\
{v}_n^{(1)}  &=& \sum_{m\neq n} \alpha_{nm}^{(1)}\,
\rho_m^{-1/2}\,{v}_m^{(0)} \,, 
\label{e:alphanm} \quad
\alpha_{nm}^{(1)}  =  \rho_m^{-1/2}
 { \left({v}_m^{(0)},\Delta_n {v}_n^{(0)}\right) \over
                   \lambda_n^{(0)} - \lambda_m^{(0)} }
\\
  \lambda_n^{(2)} &=& \sum_{m\neq n} \rho_n^{-1} \rho_m^{-1}  
                 { \left({v}_m^{(0)},\Delta_n {v}_n^{(0)}\right)^2 \over
                   \lambda_n^{(0)} - \lambda_m^{(0)} } 
                 - \rho_n^{-2}\left({v}_n^{(0)},\Delta_n {v}_n^{(0)}\right)
                   \left({v}_n^{(0)},B^{(1)} {v}_n^{(0)}\right) \,.
\ees
Also a recursion formula can be given for the higher-order
coefficients.

\subsection{Application to the perturbations $C^{(1)}$ }
Now we insert our specific problem \eq{e:A}, \eq{e:B}. 
With straightforward algebra and with a representation (for $m>n$)
\bes
    (\lambda_n^{(0)} - \lambda_m^{(0)})^{-1} \;=\;
    (\lambda_n^{(0)})^{-1}(1-\rme^{-(E_m-E_n)(t-t_0)})^{-1}
    \;=\;  (\lambda_n^{(0)})^{-1} \sum_{k=0}^\infty \rme^{-k(E_m-E_n)(t-t_0)}\,,
\ees
one finds the correction terms 
listed at the end of the first section. Initially this is so
for the first two orders, but the mentioned recursions 
allow to show that the higher orders are even more suppressed.

\subsection{Effective theory to first order}
\def\first{\mrm{1/m}}
\def\stat{\mrm{stat}}
In an effective theory, all correlation functions
\bes
  C_{ij}(t) &=& C_{ij}^{\stat}(t) \,+\, 
  \omega \,C_{ij}^{\first}(t) \,+\, \rmO(\omega^2)
\ees
are computed in an expansion in a small parameter, $\omega$,
which we consider to first order only. The notation is taken
from HQET where $\omega \propto \minv$.

We start from the GEVP in the full theory, \eq{e:gevp},
and use the form of the correction terms of the effective energies 
($t\leq2t_0$)  
\bes
    E_n^{\rm eff}(t,t_0) \;=\; \log {\lambda_n(t,t_0) \over \lambda_n(t+a,t_0)} \;=\; 
    E_n + \rmO(\rme^{-\Delta E_{N+1,n}\, t}),
\ees
see the discussion above.
Expanding this equation in $\omega$, we have 
\bes
    E_n^{\rm eff,\stat}(t,t_0) &=& a^{-1}\,\log {\lambda_n^\stat(t,t_0) \over 
                         \lambda_n^\stat(t+a,t_0)} \;=\; 
    E_n^\stat \,+\, \rmO(\rme^{-\Delta E_{N+1,n}^\stat\, t})\,,
 \\[1mm]
     E_n^{\rm eff,\first}(t,t_0) &=& {\lambda_n^\first(t,t_0) \over 
                         \lambda_n^\stat(t,t_0)} \,-\,
      {\lambda_n^\first(t+a,t_0) \over 
                         \lambda_n^\stat(t+a,t_0)} 
   \;=\; 
    E_n^\first \,+\, \rmO(t\,\rme^{-\Delta E_{N+1,n}^\stat\, t})\, .
\ees
Here $ \rmO(t\,\rme^{-E t})$ is a summary for terms 
$(b_0 + b_1 t)\rme^{-E t}$. 
As expected for first-order
perturbation theory, only the eigenvectors of the static GEVP
\bes
    C^\stat(t) \,v_n^\stat(t,t_0) \;=\; \lambda_n^\stat(t,t_0)\, C^\stat(t_0)
             \,v_n^\stat(t,t_0) \,, 
\ees
with normalization
$
      (v_m^\stat(t,t_0)\,,\,C^\stat(t_0)\,v_n^\stat(t,t_0)) =\delta_{mn}\,,
$
are needed in the formula 
\bes
  \lambda_n^\first(t,t_0) &=& \left(v_n^\stat(t,t_0)\,,\,
             [C^{\first}(t)-\lambda_n^\stat(t,t_0)C^{\first}(t_0)] 
             v_n^\stat(t,t_0)\right)
\ees
for the first-order corrections in $\omega$. 

Similarly one may expand 
\bes
   p_{01}^{\rm eff} &=& p_{01}^{\rm eff,\stat}
   + \omega \, p_{01}^{\rm eff,\first} + \rmO(\omega^2) \nonumber \\[3mm]
   p_{01}^{\rm eff,\first} &=& p_{01}^{\rm \first}
   + \rmO[\rme^{-\Delta E_{N+1,1}^\stat t_0}\,
                          \rme^{-\Delta E_{2,1}^\stat\,(t-t_0)}\,
                 (\Delta E_{N+1,1}^\first t_0 + \Delta E_{2,1}^\first\,(t-t_0))]\,
\ees
and an explicit expression for $p_{01}^{\rm eff,\first}$ is easily given.
Again it involves only the solutions of the lowest-order (in $\omega$)
GEVP, $v_n^\stat$ and $\lambda_n^\stat$, together with the first-order correlators
$C^\first$. The large energy gap $\Delta E_{N+1,1}$ controls the 
corrections.

\section{Application to static-light B$_\mrm{s}$-mesons \label{s:stat}}
We have carried out a test in quenched HQET, discretizing the static
quark by the HYP2 action and the strange quark
by the non-perturbatively $\Oa$-improved Wilson action. 
Space-time is $2L\times L^3$ with periodic boundary conditions,
$L\approx1.5\,\fm$ and we consider two lattice spacings:
$0.1\,\fm$ and $0.07\,\fm$ ($\beta=6.0219$ and $6.2885$), respectively
with $\kappa=0.133849\,,\; 0.1349798$.
The all-to-all strange-quark propagators \cite{alltoall:dublin}
are constructed
from 50 (approximate) low modes and two noise fields on each 
timeslice of 100 configurations. 

The  gauge links entering in the interpolating
fields are smeared with 3 iterations of (spatial) APE smearing
\cite{smear:ape,basak}. Then 8 different levels of Gaussian smearing 
\cite{wavef:wupp1} 
are applied
to the strange-quark field and we use a simple $\gamma_0\gamma_5$ structure
in Dirac space for all 8 interpolating fields. The local field 
(no smearing) is included to compute the decay constant.
The resulting $8\times8$ correlation function is first truncated
to an $N\times N$ one projecting with the $N$ eigenvectors 
of $C(t_i)$ with the largest eigenvalues. Here $t_i$ is taken to be
roughly 0.2 fm (i.e.\ $t_i=2a$ at $\beta=6.0219$ and $3a$ at
$\beta=6.2885$).
With $N$ not too large,
this avoids numerical instabilities and large statistical 
errors in the GEVP\cite{gevp:bern}.
We present our results for the spectrum and for the decay constant below.


\begin{figure}[htb]
\begin{center}
\hspace*{-1.2cm}
\epsfig{file=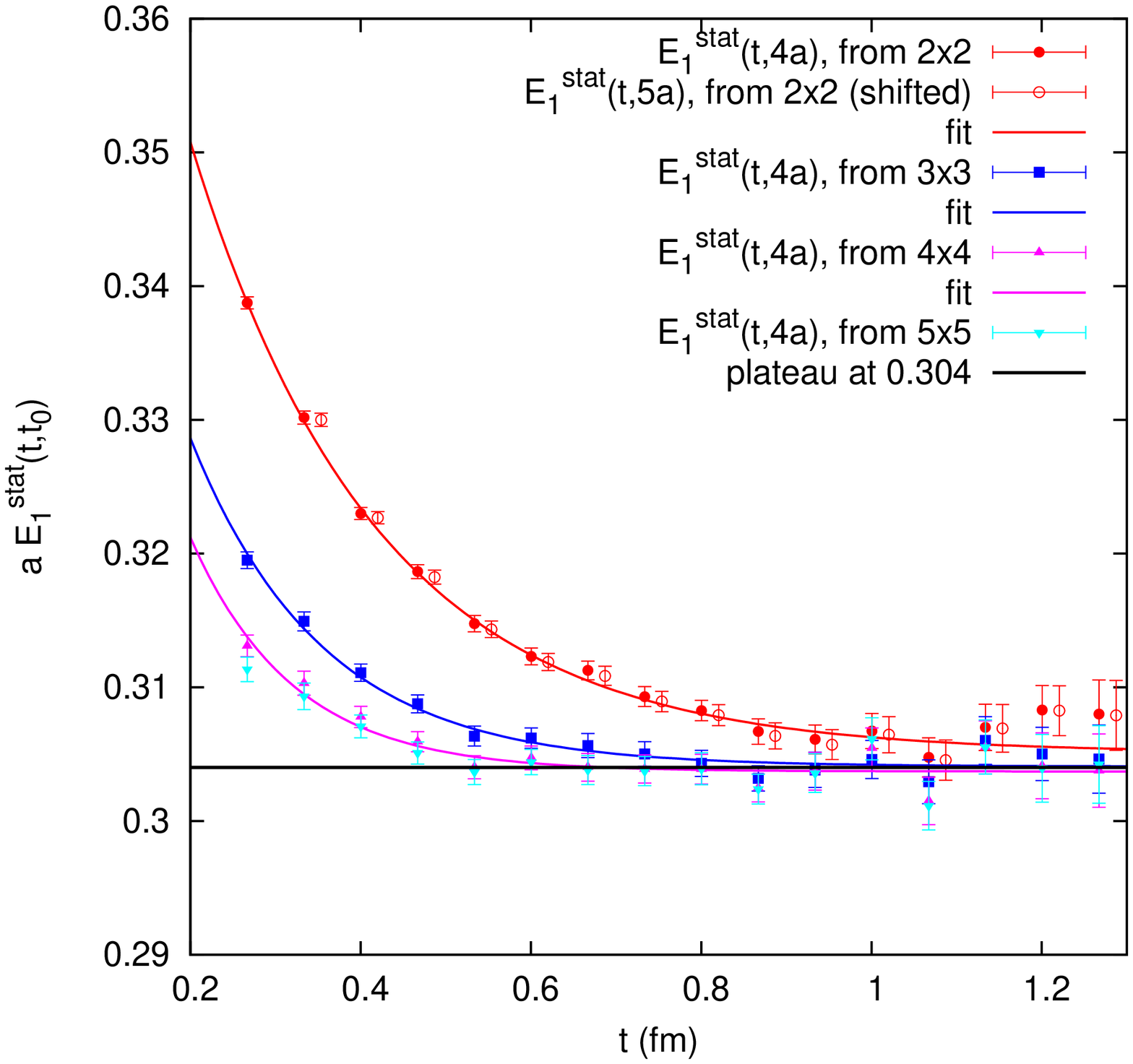,width=6.4cm,angle=270}
\hspace*{-2.2cm}
\epsfig{file=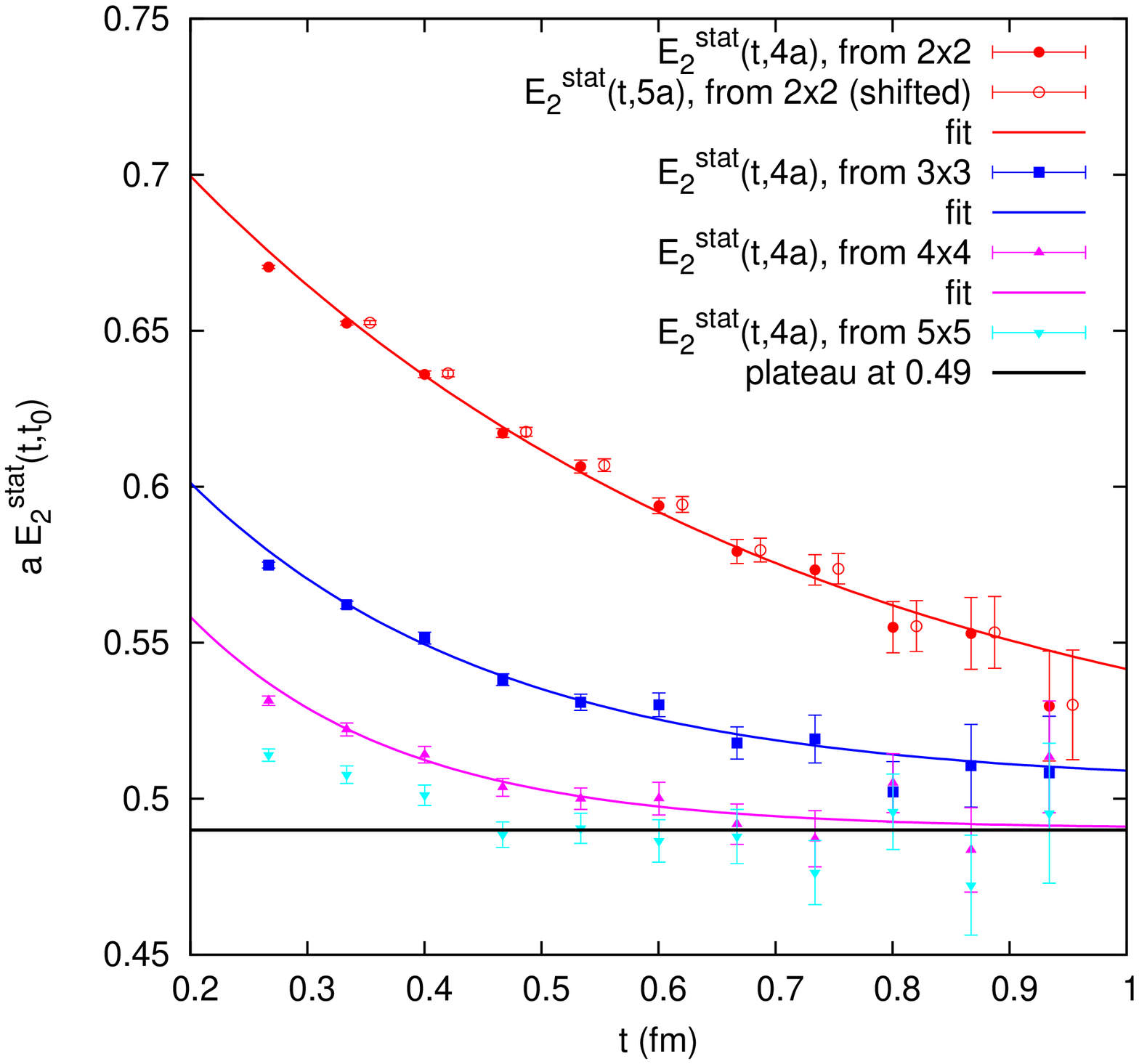,width=6.4cm,angle=270}
\hspace*{-1.cm}
\end{center}
\caption{\footnotesize{\sl The estimate $a E_n^{\rm eff,\stat}(t,t_0)$,
$n=1,2$,
as a function of $t$, for $N=2,3,4,5$ from top to bottom at $a=0.07\,\fm$.
The curves are $E_n + \alpha_N\, \rme^{-\Delta E_{N+1,1}\, t}$
(see comment about $\Delta E_{N+1,1}$ in the text).
The coefficients $\alpha_N$ are fitted for each $N$.
} }
\label{f:E163}
\end{figure}
\Fig{f:E163} shows the effective energies
\eq{e:eneff} for the lowest two levels at $a=0.07\,\fm$.
Statistical errors for the ground-state effective energy
are below a level of about $3\,\MeV$ for time
separations $t\leq1\,\fm$. Unexpectedly, these errors
are roughly independent of $t_0$ and of $N\leq5$.
The functional form of the systematic corrections
\eq{e:corren} works very well down to surprisingly small
$t$ and the independence of $t_0$ is confirmed by the
data. Since the corrections are well understood to be below
the $\MeV$--level for $t>0.6\,\fm\,, N\geq4$, we may quote
for example 
$E_1^{\stat}$ with a total error of about $1\,\MeV$.
We emphasize that what counts is of course the time separation
in physical units. The data at the coarser lattice spacing
are very similar.

For this analysis, the energy gaps on the coarser
lattice,
$a\Delta E_{N+1,1} \approx 0.46, 0.65, 0.83$, 
respectively for $N = 2, 3, 4$,
have been taken from
plateaux of $a E_n^{\rm eff,\stat}(t,t_0)$ for $N=6$. 
They have then been appropriately rescaled with the lattice spacing.
A similar procedure has been used for $a \Delta E_{N+1,2}$.


\begin{figure}[htb]
\begin{center}
\epsfig{file=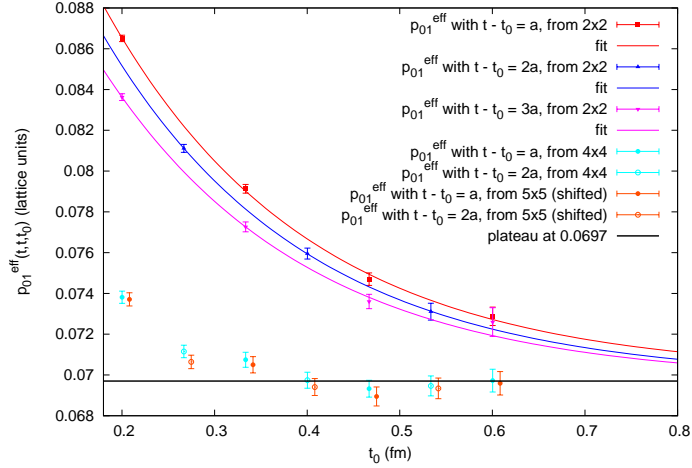,width=6.5cm,angle=270}
\end{center}
\caption{\footnotesize{\sl Bare effective static decay constant
as a function of $t_0$ for different values of $t-t_0$ at $a=0.07\,\fm$.
The curves are $F + \alpha_N\, \rme^{-\Delta E_{N+1,1}\, t_0}$
(see comment about $\Delta E_{N+1,1}$ in the text).
\label{f:psi63}
 } }
\end{figure}

\Fig{f:psi63} shows the effective decay constant, \eq{e:corren},
at the smaller 
lattice spacing. The leading corrections again dominate 
at small time already. For $N=5$ there is a rather early plateau around
$t_0=0.4\,\fm$, where both excited-state corrections are well below
the \% level and the statistical errors are around 0.7\,\%.  
The same statements hold for $a=0.10\,\fm$. Note that we fit the 
corrections separately for each $t-t_0$ and $N$ as a function of $t_0$. 
The decay of the fit parameters $\alpha_N$ as a 
function of $t-t_0$ is of the expected form \eq{e:corrp1}. 

\section{Conclusions}
From a detailed analysis of the corrections
to the eigen--values and vectors of the GEVP,
it becomes clear that $t_0$ should not be made too
small. In particular if $t_0\geq t/2$, the simple
forms \eq{e:corren}, \eq{e:corrpn} can be shown. These
corrections decay exponentially with the large gaps $E_{N+1}-E_n$.
For first-order corrections in an effective theory 
a similar suppression holds, with the energy differences of the 
lowest-order theory.  

As pointed out to us at the conference, 
the authors of \cite{gevp:dudek} studied the GEVP for a toy model with ten states
and noted that it is relevant to have $t_0$ ``large enough''.
Fig.17 of \cite{gevp:dudek} indeed
illustrates that the effective energies become independent of $t_0$ when
(roughly) $t_0 \geq t/2$ is respected.\\[1ex]

\noindent
{\bf Acknowledgements.}
This
work is supported by the  Deutsche Forschungsgemeinschaft
in the SFB/TR~09 and under grant
HE~4517/2-1, by the European community through
EU Contract No.~MRTN-CT-2006-035482, ``FLAVIAnet''.
T.M. also thanks the A. von Humboldt Foundation for support.

\bibliographystyle{JHEP}   
\bibliography{refs}           
%


\end{document}